\documentclass[12pt]{article}
 \pdfoutput=1
\usepackage{amsfonts,amsthm,amsmath,amssymb,slashed}
\usepackage[textwidth = 500 pt, textheight = 630 pt]{geometry}
\usepackage{latexsym,amsfonts,amsmath,amssymb,theorem,dsfont,wasysym}
\usepackage{amssymb}
\usepackage{amsmath}
\usepackage{amstext}
\usepackage{graphicx,epsfig}
\usepackage{epsfig}
\usepackage{verbatim} 
\usepackage{fancyhdr}
\usepackage{fancybox}
\usepackage{color}
\usepackage{ulem,bbold}
\usepackage{enumitem}
\usepackage{subfigure}
\usepackage{bbm}
\usepackage{parskip}
\usepackage[backend=bibtex, citestyle=numeric-comp, sorting=none,maxbibnames=9]{biblatex}
\usepackage{feynmf}

\definecolor{MyDarkBlue}{rgb}{0.15,0.15,0.45}
\usepackage[linktocpage=true]{hyperref}
\hypersetup{
colorlinks=true,
citecolor=MyDarkBlue,
linkcolor=MyDarkBlue,
urlcolor=MyDarkBlue,
}

\def\beq{\begin{eqnarray}}
\def\eeq{\end{eqnarray}}

\def\({\left(}
\def\){\right)}

\newcommand{\be}{\begin{equation}}
\newcommand{\ee}{\end{equation}}
\newcommand{\la}{\langle}
\newcommand{\ra}{\rangle}
\def\ea{\end{eqnarray}}
\def\ba{\begin{eqnarray}}

\def\beq{\begin{eqnarray}}
\def\eeq{\end{eqnarray}}

\def\({\left(}
\def\){\right)}

\def\la{\langle}
\def\ra{\rangle}

\def\lsim{\mathrel{\rlap{\lower3pt\hbox{\hskip0pt$\sim$}}
     \raise1pt\hbox{$<$}}}         
\def\gsim{\mathrel{\rlap{\lower4pt\hbox{\hskip1pt$\sim$}}
     \raise1pt\hbox{$>$}}}         

\def\lsim{\mathrel{\rlap{\lower3pt\hbox{\hskip0pt$\sim$}}
     \raise1pt\hbox{$<$}}}         
\def\gsim{\mathrel{\rlap{\lower4pt\hbox{\hskip1pt$\sim$}}
     \raise1pt\hbox{$>$}}}         

\usepackage{amsmath}
\usepackage{amsfonts}
\usepackage{verbatim}
\usepackage{graphicx}
\usepackage{subfigure}
\usepackage{amssymb}
\usepackage[T1]{fontenc}
\usepackage{slashed}

\begin{document}

\renewcommand{\thefootnote}{\fnsymbol{footnote}}

\makeatletter
\@addtoreset{equation}{section}
\makeatother
\renewcommand{\theequation}{\thesection.\arabic{equation}}

\rightline{}
\rightline{}


\begin{center}

{\Large \bf{Background Field Method  and Initial-Time Singularity \\ \vskip 10pt for Coherent States}}

 \vspace{1truecm}
\thispagestyle{empty} \centerline{\large  {Lasha  Berezhiani, Giordano Cintia and Michael Zantedeschi}
}

\textit{Max-Planck-Institut f\"ur Physik, \\ F\"ohringer Ring 6, 80805 M\"unchen, Germany\\}
\vskip 10pt
\textit{Arnold Sommerfeld Center, Ludwig-Maximilians-Universit\"at,\\ Theresienstra{\ss}e 37, 80333 M\"unchen, Germany}

\end{center}  
 
\begin{abstract}

The background field method is adopted for studying the dynamics of coherent states within an interacting scalar field theory. Focusing on a coherent state that corresponds to the homogeneous condensate, the quantum depletion of the expectation value of the field-operator is demonstrated to be due to the annihilation of the condensate constituents into relativistic quanta. Moreover, due to the fact that the initial field acceleration and energy for the non-squeezed coherent states are determined in terms of bare coupling constant, instead of the renormalized one, the appearance of perturbative singularities is shown to be inevitable. In other words, consistency of these states requires the finiteness of the bare coupling constant, through the resummation.

\end{abstract}

\newpage
\setcounter{page}{1}

\renewcommand{\thefootnote}{\arabic{footnote}}
\setcounter{footnote}{0}

\linespread{1.1}
\parskip 4pt

\section{Introduction}

\label{introduction}

Coherent states, parameterized by macroscopic continuous parameters, are generally deemed to be the adequate quantum counterparts to classical configurations \cite{Glauber:1963tx,Kibble:1965zza}; see also \cite{Zhang:1990fy,Zhang:1999is}. For clarity, let us kickstart the discussion by explicitly constructing the coherent state for the scalar field theory which will be the focus of the present work
\beq
|C\ra=e^{-\frac{i}{\hbar}\int {\rm d}^3 x \left( \phi_{\rm cl}(x)\hat{\pi}(x)-\pi_{\rm cl}(x)\hat{\phi}(x) \right)}|\Omega\ra\,.
\label{intro:coherent}
\eeq
The expectation value of the canonical field and its conjugate momentum, at the moment of construction of the state and in the absence of the tadpoles in the vacuum of the theory $|\Omega\ra$, can be readily obtained from canonical commutation relations 
\beq
&&\la C | \hat{\phi} |C\ra (t=0)=\phi_{\rm cl}(x)\,,\\
&&\la C | \hat{\pi} |C\ra (t=0)=\pi_{\rm cl}(x)\,.
\eeq
Therefore, as long as the functions $\phi_{\rm cl}(x)$ and $\pi_{\rm cl}(x)$ are finite in $\hbar\rightarrow 0$ limit, the state \eqref{intro:coherent} corresponds to a quantum description of the classical configuration. Equivalently, one could parameterize the coherent state by the average occupation number $N_k$ at a momentum level $k$.

Conventional methods for tackling the question of quantum corrections to the classical dynamics consist of quantizing fluctuations around the fixed classical background followed by the assessment of the semiclassical backreaction on the background; c.f. \cite{calzetta} for the overview of different methods. In this work we will discuss the advantage of constructing the state explicitly in its entirety, which gives a unique perspective on the so-called "initial-time singularity", encountered in the literature within semiclassical methods  for certain initial conditions for quantum fluctuations \cite{Cooper:1987pt,Baacke:1997zz,Boyanovsky:1998aa,Baacke:1999ia,Collins:2005cm}, in the context of coherent states.

This work has been motivated by quantum corpuscular approach to classical backgrounds developed in \cite{Dvali:2011aa,Dvali:2012en,Dvali:2012wq,Dvali:2013vxa,Dvali:2013eja,Dvali:2017eba}, where the coherent state description of dynamical systems was shown to give rise to novel quantum effects. These were demonstrated to lead to complete departure from the classical dynamics in some cases and to be of utmost importance for systems such as black holes, de Sitter and cosmic inflation. The discussion for inflation has been revisited in \cite{Berezhiani:2016grw}, and the potential ramifications for the beginning of inflation were discussed in \cite{Berezhiani:2015ola}. The question of quantum depletion of the axion condensate has been investigated in \cite{Dvali:2017ruz}. In \cite{kovtun2020breaking,kovtun2020breaking1}, quantum dynamics of condensates with a conserved charge was analyzed within 2-particle-irreducible formalism (c.f. \cite{berges2004introduction})  in the context of (1+1)-dimensional self-interacting scalar field theory, and conclusions analogous to \cite{Dvali:2017eba,Dvali:2017ruz} were drawn. The explicit dynamics of coherent states of the form \eqref{intro:coherent} were studied in \cite{Berezhiani:2020pbv} up to a certain order. The dynamics of coherent states as a quantum counterpart to classical dynamics was analyzed within quantum mechanical setting in \cite{Vachaspati:2017jtw}.

We extend the work done in \cite{Berezhiani:2020pbv} in analyzing the dynamics of \eqref{intro:coherent} to higher orders. Moreover, by adopting the well-known background field method for coherent states we avoid spurious contributions to the dynamics that were encountered in \cite{Berezhiani:2020pbv}. Our goal, in this article, is to capture the results of the S-matrix analysis of \cite{Dvali:2017eba} within our direct computation of the real-time dynamics of coherent states. Following \cite{Dvali:2017eba}, in a weakly interacting theory the homogeneously oscillating background with frequency $\omega$ corresponds to a quantum state with large occupation number $N$ of quanta with the same frequency. Although the classical description is expected to hold for a large period of time for such configurations, the scattering among quantum constituents could lead to the gradual breakdown of the semiclassical approximation. The timescale after which the quantum dynamics is expected to deviate from its classical counterpart significantly was coined as \textit{quantum break-time} \cite{Dvali:2017eba} and was found to be given by\footnote{For systems exhibiting semiclassical instability the quantum break-time has been shown to scale as  $t_{\rm qb} \sim \gamma^{-1} \log{N}$ in \cite{Dvali:2013vxa}, where $\gamma$ is the Lyapunov exponent of the associated instability; see also \cite{kovtun2020breaking1} for the derivation within 2-particle-irreducible formalism.
Clearly, a system undergoing quantum breaking is also necessairly scrambling information \cite{Dvali:2013vxa}. Black holes, for example, are believed to be fast scramblers \cite{hayden2007black}, i.e. they thermalize in the above mentioned logarithmically dependent timescale ($\gamma^{-1}\sim R_{\rm g},$ $N=S$, $S$ being its entropy). This is due to the excitation of their quasi-normal modes (Lyapunov exponents) under external perturbations. In this optic, the scrambling time corresponds to the time needed by the black hole to adjust to the received information.
}

\beq
\label{eq:powerlawqb}
    t_{\rm qb} \sim \frac{1}{\lambda^\alpha \left(\lambda N\right)^{\beta}\omega};
\eeq
where $\lambda$ is a dimensionless coupling constant and $\alpha,\beta\geq 1$ are determined by the scattering channel dominating the depletion. We will show that for a massive scalar field with quartic self-interaction the depletion of the 1-point expectation value of the field in a coherent state \eqref{intro:coherent}, with $\phi_{\rm cl}(x)=const$ and $\pi_{\rm cl}(x)=0$, is dominated by $4\rightarrow 2$ annihilation of the constituents for $\lambda N\lesssim 1$, which is the lowest order kinematically allowed depletion channel for the theory at hand. The corresponding timescale is given by \eqref{eq:powerlawqb} with $\alpha=1$ and $\beta=3$, reproducing the S-matrix result of \cite{Dvali:2017ruz}.

The article is organized as follows. In Sec. 2, relevant aspects of the background field method are reviewed. In Sec. 3, this technique is applied to coherent states and the 1-loop quantum dynamics of the expectation value of the field  is analyzed both analytically and numerically, in a coherent state corresponding to the homogeneous condensate. In Sec. 4, the depletion of the coherent state due to particle production is discussed and the connection with the S-matrix processes for the annihilation of the condensate constituents is made. In Sec. 5, certain aspects of 2-loop corrections and their implications for the construction of the coherent state are discussed. In Sec. 6, the appearance and implications of the initial-time singularity in the dynamics of coherent states are discussed. Sec. 7 is devoted to the summary and outlook.

\section{Background-Field Method}

We begin by overviewing the background-field method for studying quantum dynamics, see e.g. \cite{Baacke:1997zz}. This is an important step necessary to compare standard results found in the literature to what we will derive using a coherent state description. In particular, even though the two descriptions are in one to one correspondence and both could describe exactly the quantum dynamics of a given system, in the latter we have full control over the state of the system itself. 

In order to underline the differences between various approximations we will refer to throughout the paper, we begin from the discussion of the classical dynamics and then progress towards the fully fledged quantum dynamics described by the background-field method that is in principle exact, but requires an $\hbar$ expansion, for practical purposes.
For definiteness, we will focus on a massive scalar field with quartic self-interaction with Lagrangian: 
\beq
\mathcal{L}=-\frac{1}{2}(\partial_\mu\hat{\phi})^2-\frac{1}{2}m^2 \hat{\phi
}^2-\frac{\lambda}{4!}\,\hat{\phi}^4\,.
\label{eq:Lagr}
\eeq
\subsection*{Classical and semi-classical dynamics }
The classical analogue of Lagrangian \eqref{eq:Lagr} is easily obtained by replacing the field operator $\hat{\phi}$ with a c-number field $\phi$.
The classical dynamics follows from the equation of motion
\beq
\left(-\Box+m^2\right)\phi+\frac{\lambda}{3!}\,\phi^3=0.
\label{eq:eom}
\eeq
Being classical, this equation describes the dynamics in $\hbar\rightarrow 0$ limit.

Now, in order to capture some of the quantum effects, it suffices to go beyond the classical approximation by quantising fluctuations around the classical configuration. This is done by quantising the linear equations for fluctuations
\beq
\left( -\Box+m^2+\frac{\lambda}{2}\Phi_{\rm cl}^2 \right)\hat{\psi}=0\,,
\eeq
where $\Phi_{\rm cl}$ is the solution of classical equations \eqref{eq:eom} and  the fluctuation operator has been defined by means of the following decomposition of the field operator $\hat{\phi}=\Phi_{\rm cl}+\hat{\psi}$.

Although this approximation corresponds to keeping $\hbar\ll 1$ but finite, by ignoring quantum back-reaction on the background, it assumes $\Phi_{\rm cl}\gg \sqrt{\la\hat{\psi}\hat{\psi}\ra}$. 

The important, usually overlooked, point stressed in a series of papers \cite{Dvali:2011aa,Dvali:2012en,Dvali:2012wq,Dvali:2013vxa,Dvali:2013eja,Dvali:2017eba}, is that the results obtained in this way are exact only in $\Phi_{\rm cl}\rightarrow \infty$ limit. Consequently, one should be extremely cautious when attributing a physical meaning to the semi-classical back-reaction from quantum fluctuations on the background itself.

\subsection*{Full-quantum dynamics}
\label{backgroundmethod}

The background-field method adopted in Ref.~\cite{Baacke:1997zz} for computing the quantum corrections to the classical dynamics consists of analysing the coupled system of equations of the one-point expectation value in a quantum state and the equation for the correlation functions for fluctuations, with the latter quantised as the deviation from the one-point function. The starting point for the derivation is Heisenberg's operator equation
\beq
\left( -\Box+m^2 \right)\hat{\phi}+\frac{\lambda}{3!}\,\hat{\phi}^3=0\,.
\label{hamiltoneq}
\eeq

For the system in a quantum state $|\Psi\ra$, one performs the following decomposition of the Heisenberg picture operator
\beq
\hat{\phi}=\Phi+\hat{\psi}\,,\qquad \text{with}\qquad \Phi\equiv \la\Psi|\hat{\phi}|\Psi\ra\,.
\label{eq:decomp}
\eeq
Notice that $\la\Psi|\hat{\psi}|\Psi\ra=0$ by definition. As a result, the expectation value of Eq.~\eqref{hamiltoneq} reduces to
\beq
\label{bckgreq}
\left( -\Box+m^2 +\frac{\lambda}{2}\la \Psi | \hat{\psi}^2(x,t) |\Psi\ra \right)\Phi(x,t)+\frac{\lambda}{3!}\,\Phi^3(x,t)+\frac{\lambda}{3!}\la\Psi | \hat{\psi}^3(x,t) | \Psi\ra=0\,.
\eeq
The substitution of the above decomposition into \eqref{hamiltoneq} and subtracting \eqref{bckgreq} from it yields
\beq
\left( -\Box+m^2 +\frac{\lambda}{2}\,\Phi^2(x,t) \right)\hat{\psi}(x,t)+\frac{\lambda}{2}\Phi(x,t)\left( \hat{\psi}^2(x,t)-\la\Psi | \hat{\psi}^2(x,t) | \Psi\ra\right)\nonumber \\
+\frac{\lambda}{3!}\left( \hat{\psi}^3(x,t)-\la\Psi | \hat{\psi}^3(x,t) | \Psi\ra\right)=0\,.
\label{psieq}
\eeq
The latter equation can be straightforwardly converted into the set of equations for $n$-point correlation functions if one multiplies it by  the corresponding number of operators and evaluates everything over $|\Psi\rangle$. For example, the 2-point function satisfies
\beq
\label{2pointpsi}
\left( -\Box^{(x,t)}+m^2 +\frac{\lambda}{2}\,\Phi^2(x,t) \right)\la\Psi|\hat{\psi}(x,t)\hat{\psi}(x',t')|\Psi\ra+\frac{\lambda}{2}\,\Phi(x,t) \la\Psi | \hat{\psi}^2(x,t)\hat{\psi}(x',t') | \Psi\ra\nonumber\\
+\frac{\lambda}{3!}\la\Psi | \hat{\psi}^3(x,t)\hat{\psi}(x',t') | \Psi\ra=0\,.
\eeq
We would like to stress that the equations \eqref{bckgreq}-\eqref{2pointpsi} are exact and, if solved consistently, would provide a consistent evolution of the 1-point as well as higher-order correlation functions. This is a direct consequence of the fact that the decomposition \eqref{eq:decomp} is exact, differently from the one adopted in the semi-classical approximation. Obviously, one needs to perform some kind of perturbative expansion in order to proceed. The most natural one being the $\hbar$ expansion, i.e. a loop expansion. 

Up to this point the discussion has been very general, we have not even specified the quantum state. The analysis simplifies if the state in question has macroscopic 1-point function, i.e. $\lim_{\hbar\rightarrow 0}\Phi\neq 0$. In this case, if one is interested in leading order $\hbar$ corrections to the evolution of $\Phi$ it suffices to solve Eq.~\eqref{bckgreq} without the last term. For this task one would need to solve the 2-point function equation~\eqref{2pointpsi} at tree level, that is without the last two terms. Even at this level, the actual procedure is nontrivial as the solution is a complicated function of the classical coupling $\lambda$. However, it can be solved numerically for homogeneous $\Phi(t)$ (with suitable initial conditions) to a desirable order in coupling, as it was done in \cite{Baacke:1997zz}. More generally, if we would like to know the evolution of higher order correlators we need to know initial conditions for all of them. The specification of state $|\Psi\ra$ corresponds precisely to this: the notion of the initial conditions. We will see below how all this comes about within the coherent state description of the system.

Let us conclude this section by pointing out that, when one tackles the question of quantum evolution of a classical field configuration of interest, it is usually attempted to guess the initial conditions for the correlators. This is equivalent to writing down the quantum state of the system based on a 1-point expectation value, which could lead to misleading conclusions. It goes without saying that this comment only applies to the case when the only input for the quantum computation is taken to be the classical properties of the system. There are cases when relevant initial conditions are known for mode functions (i.e. for the correlation functions) as well. The examples are the systems initially prepared in thermal equilibrium with an external medium; for instance, if one is interested in the quantum evolution of a Bose-Einstein condensate which was brought to zero temperature at some initial time, then the initial state can be justifiably taken to be the vacuum of the Bogoliubov modes.

\section{Coherent State Description}

The background-field method outlined in the previous section, as powerful as it can be, requires an input in the form of initial conditions for all the correlation functions in order to find the evolution of the system fully. Moreover, the choice of these conditions corresponding to a consistent quantum state may be challenging. Not to mention that it may be impractical for calculating certain quantum information, i.e. the quantum entanglement generated as a result of the evolution. These in general would require the computation of high-order correlators, or alternatively the evolution of the state itself. The importance of understanding the latter has been underlined for various systems in Refs.~\cite{Dvali:2011aa,Dvali:2012en,Dvali:2012wq,Dvali:2013vxa,Dvali:2013eja,Dvali:2017eba}.

Therefore, it is important to take a step back and discuss the procedure for constructing a state itself. The systems possessing an approximate classical description, are proven to be adequately described in terms of coherent states. Following Ref.~\cite{Berezhiani:2020pbv}, we construct the coherent state at initial time $t=0$ out of the vacuum state of the interacting quantum field theory at hand and the canonical degrees of freedom as
\beq
|C\ra=e^{-i\int {\rm d}^3 x \left( \phi_{\rm cl}(x)\hat{\pi}(x)-\pi_{\rm cl}(x)\hat{\phi}(x) \right)}|\Omega\ra\,,
\label{eq:coherent}
\eeq
with $|\Omega\ra$ denoting the Hamiltonian eigenstate with the lowest possible energy eigenvalue, $(\hat{\phi},\hat{\pi})$  satisfying canonical commutation relations and $(\phi_{\rm cl},\pi_{\rm cl})$ being c-number functions.
The convenient property of this state is that it has unit norm and satisfies the following to all orders in $\lambda$ at the initial moment of time $t=0$
\beq
&&\la C | \hat{\phi} |C\ra (t=0)=\phi_{\rm cl}(x)\,,\\
&&\la C | \hat{\pi} |C\ra (t=0)=\pi_{\rm cl}(x)\,.
\eeq
These relations determine the initial conditions for the 1-point expectation values. Since the state has been formulated in its entirety, we can fish out the initial conditions for other correlators as well. For instance, we have the following for the 2-point functions
\beq
\label{init2point}
&&\la C | \hat{\phi}(x,0)\hat{\phi}(y,0) |C\ra=\phi_{\rm cl}(x)\phi_{\rm cl}(y)+\la \Omega | \hat{\phi}(x,0)\hat{\phi}(y,0) |\Omega\ra\,,\\
&&\la C | \hat{\pi}(x,0)\hat{\pi}(y,0) |C\ra=\pi_{\rm cl}(x)\pi_{\rm cl}(y)+\la \Omega | \hat{\pi}(x,0)\hat{\pi}(y,0) |\Omega\ra\,.
\eeq
In other words, if we were to define the fluctuation operator as $\hat{\psi}\equiv \hat{\phi}-\la C|\hat{\phi}|C\ra$, similar to the previous section, we would find its initial 2-point function to be given by the vacuum correlator. 

Instead of solving the coupled system of equations outlined in the previous section as a consistent implementation of the background-field method, we stick with the explicit representation of the quantum contribution to the equation of the 1-point function in terms of the coherent state. We begin from Eq.~\eqref{bckgreq} and take into account that the last term starts to contribute only at $\hbar^2$ order (i.e. at 2-loop). We further rewrite the third term in parenthesis in terms of the un-decomposed operator
\beq
\left( -\Box+m^2 \right)\Phi(x,t)+\frac{\lambda}{3!}\,\Phi^3(x,t)+\frac{\lambda}{2}\,\Phi(x,t)\left[\la C | \hat{\phi}^2(x,t) |C\ra-\la C | \hat{\phi}(x,t) |C\ra^2 \right]+\mathcal{O}(\hbar^2)=0\,,
\label{cohst1pointeq}
\eeq
with $\Phi(x,t)\equiv \la C | \hat{\phi}(x,t) |C\ra$ as before. Here, we have borrowed a trick from the background-field method in order to drop the manifestly 2-loop order contribution. 
The second term in square brackets is obviously $\Phi^2$, however we keep it in the given form to underline that we will be evaluating the bracketed expression explicitly up to a desired order in coupling constant. We would like to stress that the only approximation made at this level is the loop expansion and Eq.~\eqref{cohst1pointeq} is exact at $\mathcal{O}(\hbar)$. It is important to keep in mind that when evaluating the bracketed term we will encounter higher loop contributions that would need to be dropped for the consistency of the computation.

Throughout this work we will be focussing on the idealised coherent state with $\phi_{\rm cl}(x)=\phi_0={\rm const}$ and $\pi_{\rm cl}(x)=0$. The direct consequence of this homogeneity and the translation invariance of the Hamiltonian is the homogeneity of the 1-point expectation values, which will serve as a simplifying factor; e.g. $\Phi(t)\equiv \la C |\hat{\phi}|C \ra$ remains homogeneous at all times for homogeneous coherent states.

\begin{figure} 
	\centering
	\includegraphics[scale=0.43]{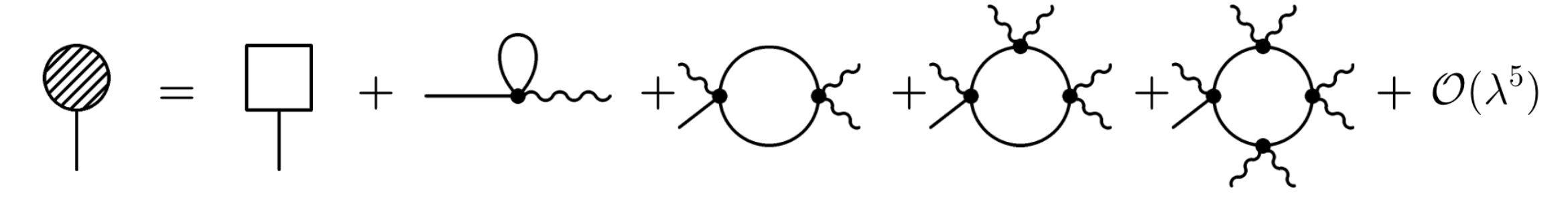} 
	\caption{The irreducible diagrams that contribute to the one-point expectation value of $\phi$ (indicated on the l.h.s. as the full circle), up to $\lambda^4$. Here, the box stands for action variation $\delta S/ \delta \phi|_{\phi=\Phi(t)}$ and the wiggled lines correspond to $\Phi(t)$ insertions. The vertices are clockwise time-ordered. The second (third) term on the r.h.s. are divergent and are responsible for mass (coupling) renormalization. 
	}
	\label{fig:diag}
\end{figure}

The direct evaluation of the 1-point expectation value $\Phi(t)$ involves two types of contributions: 1-particle irreducible (1PI) and reducible ones. The first being encoded in the diagrams of Fig.~\ref{fig:diag}, while the latter may be understood as an irreducible diagram where some wiggled lines are replaced by propagators connecting irreducible sub-diagrams. One of the important points we would like to emphasize is that the utilization of Eq.~\eqref{cohst1pointeq} resums the reducible contributions into $\Phi$. This is a direct consequence of the fact that the reducible contributions cancel within the bracketed expression. Therefore, if one reads the wiggled lines of Fig.~\ref{fig:diag} as $\Phi(t)$ up to a certain necessary order in coupling (and not as the free solution $\Phi_0(t)=\phi_0 {\rm cos}(mt)$), reducible diagrams are automatically taken into account  and may be dropped in the calculation. This is carefully shown in appendix A and we redirect the reader there for more technical details on this point. 
Finally, it appears as if 1PI terms could be re-summed too into a loop of the propagator with a shifted mass (although a time-dependent one). The caveat impeding such a simplification will be discussed at the end of this section.

Now, by direct evaluation of the expectation values up to order $\lambda^4$ , Eq.~\eqref{cohst1pointeq} can be brought to the following form
\begin{flalign}
\left( \partial_t^2+m_{\rm ph}^2 \right)\Phi(t)+\frac{\lambda}{3!}\,\Phi^3(t)=&\frac{\lambda^2}{2}\Phi(t)\int_0^t {\rm d}t_1\Phi^2(t_1) S_1(t,t_1)\nonumber\\&-\frac{\lambda^3}{2}\Phi(t)\int_0^{t} {\rm d}t_1\int_{0}^{t_1} {\rm d}t_2 \Phi^2(t_1)\Phi^2(t_2)S_2(t,t_1,t_2)\nonumber\\&+\frac{\lambda^4}{8}\Phi(t)\int_0^t {\rm d}t_1\int_0^{t_1}{\rm d}t_2\int_0^{t_2}{\rm d}t_3\Phi^2(t_1)\Phi^2(t_2)\Phi^2(t_3) S_3(t,t_1,t_2,t_3)\,
\label{eq:1loop}
\end{flalign}
where:
\begin{flalign}
&S_1(t,t_1)=\int\frac{{\rm d}^3p}{(2\pi)^3(2E_{\rm p})^2}\sin\left(2E_{\rm p}(t-t_1)\right)\\
&S_2(t,t_1,t_2)=\int\frac{{\rm d}^3p}{(2\pi)^3(2E_{\rm p})^3}\Big(\cos\left(2E_{\rm p}(t_1-t_2)\right)-\cos\left(2E_{\rm p}(t-t_1)\right)\Big)\\
&S_3(t,t_1,t_2,t_3)=\int\frac{{\rm d}^3p}{(2\pi)^3(2E_{\rm p})^4}\Big(\sin \left(2E_{\rm p}(t-t_3)\right)-\sin \left(2E_{\rm p}(t_1-t_3)\right)\nonumber\\&\hskip 170pt+(\sin\left(2E_{\rm p}(t-t_1)\right)\cos\left(2E_{\rm p}(t_2-t_3)\right)\Big)
\end{flalign}
The procedure to evaluate $\langle C|\phi|C\rangle$ and $\langle C|\phi^2|C\rangle$ is shown in Appendix A. Here, we dropped the gradient due to the homogeneity of the 1-point function. We have also performed the mass renormalization, removing one of the manifest divergencies, by
\beq
m_{\text{ph}}^2\equiv m^2+\frac{\lambda}{2}\langle\hat{\phi}^2\rangle\,.
\label{eq:1loopmass}
\eeq
The remaining divergence resides in the first term on the right hand side of Eq.~\eqref{eq:1loop} which can be taken care of by the coupling renormalization in analogy with \cite{Berezhiani:2020pbv}. Namely, the UV divergence can be isolated by performing time-integration by parts which can be subsequently absorbed by the third term on the left hand side via coupling renormalization
\begin{flalign}
\lambda_{\text{ph}}\equiv\lambda-3\lambda^2\int \frac{{\rm d}^3p}{(2\pi)^3(2E_{\rm p})^3}\,.
\label{eq:1loopcoupling}
\end{flalign}
As one should have expected, the adopted renormalization prescription is identical to its S-matrix counterpart. However, once condition \eqref{eq:1loopcoupling} is imposed in equation \eqref{eq:1loop}, a time-dependent divergence is generated. This is known in literature as 'initial time singularity' \cite{Baacke:1999ia} 
because it diverges only for $t=0$. Although it is singular at the level of the equation of motion, it gives a regular contribution to 1-point function upon integration. The issue of initial-time singularities based on the choice of the initial state is discussed in Sec.~\ref{sec:initial}.

Notice that we can simply substitute the classical counterpart of $\Phi$ in terms appearing on the right hand side of \eqref{eq:1loop}, as they are manifestly quantum due to loop integrals. Working at the level of the equation of motion, rather than computing the 1-point expectation value, leads to a much more straightforward calculation: evaluating the difference in Eq.~\eqref{cohst1pointeq} requires  studying the 1-point and 2-point functions to a lower order in $\lambda$ than the one required for a direct computation of the 1-point expectation value, as shown in Appendix A. In other words, we trade an explicit calculation of the 1-point function up to $\lambda^4$ with an analogous calculation of the 1-point and 2-point functions up to $\lambda^{3}$. 
 Therefore, one less time integral is involved in the terms appearing in the equation of motion as compared to the one-point expectation value.

The perturbative solution to \eqref{eq:1loop} can be obtained iteratively in $\lambda$, resulting in a generalization of the result of \cite{Berezhiani:2020pbv} to include  up-to order $\lambda^4$ corrections. The result can be found in appendix A, together with a complementary/direct derivation. As it is well known, a blatant perturbative analysis leads to (spurious) secular instabilities in nonlinear systems \cite{berges2004introduction} and a subtle  handling is required to extract physical trends. This can be achieved by mixing different orders in a way that removes unphysical behaviour. Namely, instead of approximating $\Phi$'s appearing in the interaction terms of \eqref{eq:1loop} by its perturbative counterpart evaluated to a minimal required order, we will treat them fully as an unknown function and solve the equation for it numerically. The outcome of this computation will be presented in the next section, followed by a physical discussion.

We would like to conclude this section by touching upon the fact that the structure of the diagrams of Fig.~\ref{fig:diag} resembles a geometric series that resums into a time-dependent mass-shift. For clarity, let us recall the 1-loop equation of motion for $\Phi(t)$ in the compact form of the previous section
\beq
\left( \partial_t^2+m^2  \right)\Phi(t)+\frac{\lambda}{3!}\Phi^3=-\frac{\lambda}{2}\la C | \hat{\psi}^2(x,t) |C\ra\Phi_{\rm cl}(t)\,.
\label{eq1point}
\eeq
Here $\Phi_{\rm cl}$ stands for the solution to classical equations of motion. Notice that the right hand side of \eqref{eq1point} is manifestly quantum and vanishes in the classical limit. Its evaluation at 1-loop requires the knowledge of the tree-level dynamics of $\hat{\psi}\equiv \la C| \hat{\phi}  | C\ra-\Phi(t)$, which is readily given by
\beq
\left( -\Box+m^2 +\frac{\lambda}{2}\,\Phi_{\rm cl}^2(t) \right)\hat{\psi}(x,t)=0\,.
\label{psieq}
\eeq
Let us reiterate that we have assumed the homogeneity of the 1-point function, which is a direct consequence of the translation invariance of the initial coherent state in question and of the Hamiltonian. We solve \eqref{psieq} following the standard procedure, see e.g. \cite{Baacke:1997zz}. One begins with the mode decomposition
\begin{equation}
    \hat{\psi}(x) = \int \frac{{\rm d}^3{k}}{(2\pi)^3}\frac{1}{\sqrt{2 \omega_{k}^0}}\left[a_{k} U_{k}(t) + a_{-{k}}^\dagger U_{-{k}}^*(t)\right] e^{i \vec{k}\cdot \vec{x}},
\end{equation}
with initial conditions $U_{{k}}(0)=0$ and $\dot{U}_{{k}}(0) = -i\, \omega_{{k}}^0$. The choice of $\omega_{{k}}^0$ is dictated by the quantum state of interest, which sets initial conditions for correlation functions. Namely, for unsqueezed coherent states \eqref{eq:coherent} the initial condition for 2-point function \eqref{init2point} corresponds to $\omega_{{k}}^0=\sqrt{{k}^2+m^2}$ at the tree-level.

As a result of this decomposition the relevant set of equations governing the one-loop dynamics of $\Phi$ take the following form
\beq
\label{eq:1loopeom}
&&  \left( \partial_t^2+m^2 \right)\Phi(t)+\frac{\lambda}{3!}\Phi^3(t)=-\frac{\lambda}{2}\Phi_{\rm cl}(t)\int \frac{{\rm d}^3{k}}{(2\pi)^3}\frac{1}{2\omega_{{k}}^0}\left|U_{{k}}(t)\right|^2\,,\\
  && \left(\partial_t^2+ \omega_{{k}}^2 \right)U_{{k}}(t)=0\,,
  \label{eq:ukevol}
\eeq
with $\omega_{{k}}= \sqrt{{k}^2+m^2 + \lambda/2\, \Phi_{\rm cl}(t)^2}$. It is rather straightforward to show how the terms in Eq.~\eqref{eq:1loop} are recovered from Eq.~\eqref{eq:1loopeom}. Indeed it is sufficient to expand Eq.~\eqref{eq:ukevol} perturbatively w.r.t. coupling constant. Such a formulation makes the diagramatic structure of Fig.~\ref{fig:diag} transparent. Furthermore, in this form the 1-loop dynamics of the 1-point function can be obtained numerically without even resorting to the expansion in $\lambda$, which is precisely what was done in \cite{baacke1997nonequilibrium}.

\subsection{Analysis}\label{sec:numerics}

In this subsection Eq.~\eqref{eq:1loop} is evaluated numerically. Let us stress that the 1-loop dynamics encoded by the system of Eqs.~\eqref{eq:1loopeom}-\eqref{eq:ukevol} has already been simulated in Ref.~\cite{baacke1997nonequilibrium} for $\lambda/(8 \pi^2)=0.1$ and an initial value of the field (in mass units) of $\Phi(0)=5$. Even though this choice of parameters is still within the validity of the 1-loop expansion, in the aforementioned numerical analysis an initial strong damping takes place as the field, within 5 oscillations, is halved in amplitude. In fact, we also checked that for their choice of initial conditions, the system is within a  parametrically resonating instability band. Therefore, the exponentially fast growing modes lead to an initial strong damping of the mean field dynamics after which the system starts depleting according to the well known scaling $t^{-1/3}$ \cite{boyanovsky1966new}.

Since we would like to verify whether the perturbatively expanded Eq.~\eqref{eq:1loop} and the full 1-loop dynamics of Eq.~\eqref{eq:1loopeom} have a similar leading behaviour (at least for small time), we focus on a region away from the parametrically unstable modes, with collective coupling $\lambda \Phi^2(0)$ close to unity\footnote{The coupling chosen by Ref.~\cite{baacke1997nonequilibrium} invalidates the perturbative $\lambda$ expansion in Eq.~\eqref{eq:1loop}.}. We therefore focus on the case $\Phi(0)=5$ and $\lambda = 0.1$. Moreover, being on the lattice, we simulate all the equations of motion in terms of unrenormalized quantities, as a natural renormalization prescription is given by the lattice finiteness. The UV and IR cutoff were chosen so that the former (latter) is much higher (lower) than the physical frequencies of the system. 
\begin{figure}[t]
    \centering
        \includegraphics[width=.8\textwidth]{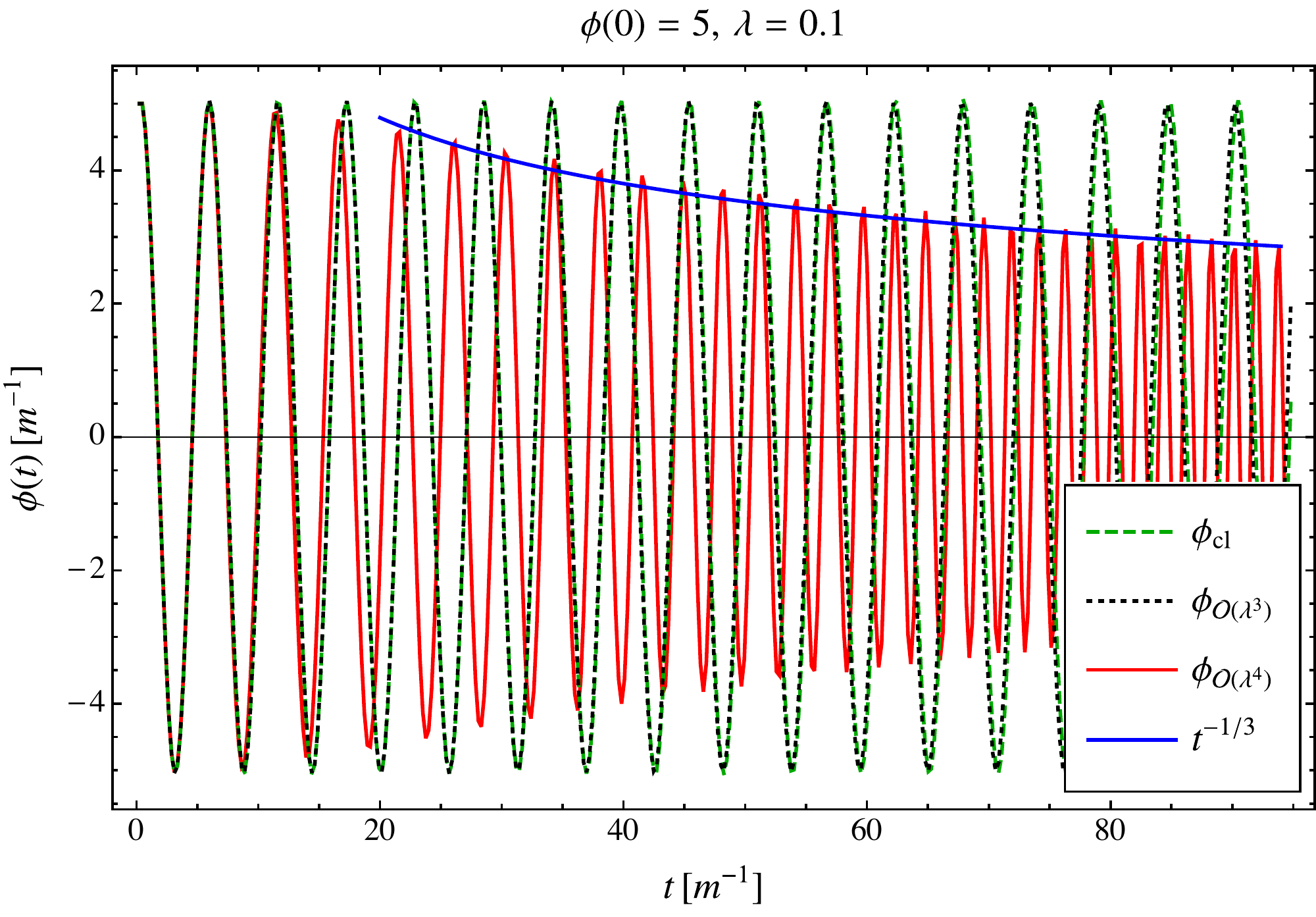}
        \caption{Simulations of Eq.~\eqref{eq:1loop} including different $\mathcal{O}(\lambda)$ terms. The green line corresponds the the classical solution of the equation of motion while the black line to the one-point function up to $\lambda^3$ and $\hbar$ corrections. No qualitative deviations are seen between those two lines. A completely different behaviour is obtained including the $\lambda^4$ terms in the one-point function (red line). At this order, the $4\rightarrow 2$ annihilation channel of constituent quanta opens up and the condensate starts depleting. The blue line represents the scaling $t^{-1/3}$ usually obtained in literature from the loop-analysis: as we may see this matches the scaling of depletion obtained from the $\lambda^4$ perturbative analysis.}
        \label{fig:pert}
\end{figure}
Numerical simulation of Eq.~\eqref{eq:1loop} is shown in Fig.~\ref{fig:pert} \footnote{Each higher $\lambda$ order in Eq.~\eqref{eq:1loop}, involves a new time integration therefore undermining both the speed of the simulation, as well as its rate of convergence. To obviate the problem we evaluated the integrals of \eqref{eq:1loop} via Monte-Carlo with Sobol quasi-random sequences. Indeed more advanced algorithms such as VEGAS \cite{lepage1978new} could easily improve the convergence of the simulation, but for our current purpose, and result, the aboce mentioned method proved sufficient.}. As it can be seen, no qualitative deviations are seen between the classical non-linear dynamics and the perturbed quantum one up to $\mathcal{O}(\lambda^3)$ (small deviations are not visible due to the plot resolution). 
The situation is dramatically different at order $\lambda^4$,  which is precisely the order at which $4\rightarrow2$ ($4$ constituent quanta into $2$ fluctuations, c.f. Fig.~\ref{fig:fig2}) annihilation channel opens up. The mean field dissipates as  $t^{-1/3}$ as seen in Fig.~\ref{fig:pert}.
Such a scaling is well known given the full 1-loop equation \eqref{eq:1loopeom} \cite{boyanovsky1966new}. Interestingly, we showed here that it is recovered by simply including terms of high enough order in coupling $\lambda$. Indeed, we verified that the dynamics obtained perturbatively in coupling and the one obtained from the full 1-loop analysis overlap for the timescale shown in Fig.~\ref{fig:pert} \footnote{For longer timescales the memory integrals appearing in \eqref{eq:1loop} are no longer numerically reliable and therefore no statement can be made regarding the late time behaviour. After a certain time new diagrams become relevant for the dynamics (e.g. the ones responsible for rescattering of excitations and thermalization \cite{berges2004introduction}).}. By no means is the overlap between two expansions trivial. In fact, for the parameters at hand one should have naively expected the significance of semi-classical nonlinear corrections  not accounted for by our $\mathcal{O}(\lambda^4,\hbar)$ computation. However, it seems that higher order channels describing the annihilation of more than four constituents are dynamically negligible. Such contributions  are in fact accounted for within full 1-loop treatment.

\section{Quantum Depletion}

In this section, we would like to analogize the semi-classical dynamics of the coherent state, laid out above, to the S-matrix description of quantum depletion of classical systems \cite{Dvali:2011aa,Dvali:2012en,Dvali:2012wq,Dvali:2013eja,Dvali:2013vxa,Dvali:2017eba,Dvali:2017ruz}. The coherent state we have focused on corresponds to the homogeneous initial field displacement $\phi_0$. Classically, such a field configuration would result into anharmonic oscillations, with characteristic frequency depending on $\phi_0$. For small amplitudes, i.e. $\lambda \phi_0^2\ll m^2$, the classical oscillation takes place with frequency $m$ and nonlinearities play insignificant role in the dynamics for an extended period of time. However, the small correction to the oscillation frequency $\delta \omega\sim \lambda\phi_0^2/m$ leads to the cumulative offset of the phase relative to non-interacting case, which becomes significant over  the classical timescale $t_{\rm cl}\sim \delta\omega^{-1}$. When necessary, one can correct this offset by shifting the phase of the free solution extending the consistency for another $t_{\rm cl}$. Moreover, due to periodicity, the departure in question is obviously a transient effect. For large amplitudes, on the other hand, the characteristic frequency becomes of order $\sqrt{\lambda}\phi_0$. In the latter case, the S-matrix analysis is complicated by the fact that the coherent state at hand cannot be viewed as a condensate of on-shell particles, instead one should revisit the notion of particle and consider them to be significantly off-shell with effective mass set by the characteristic classical frequency. In the former case, on the other hand, one can proceed to analyze the quantum dynamics as the S-matrix process for the condensate of on-shell particles,  due to the fact that \cite{Dvali:2017eba}
\beq
   \label{eq:coherentev}
\lim_{\hbar\rightarrow 0} \Phi(t)= \langle C| \hat{\phi}|C\rangle(t)\simeq \phi_0 \cos \left(m\, t \right)\,,\qquad {\rm for }\qquad \lambda\phi_0^2\ll m^2\,
\eeq
(keeping in mind the above-mentioned cumulative phase-shift)

Notice that, since we have been working with infinite homogeneity, the regulating volume must be taken to be larger than any other length-scale in the problem. Therefore, we are dealing with a state of macroscopic number of particles, populated by zero-momentum particles with number density $n\sim m\phi_0^2$. Denoting the total particle number by $N$, one can study the S-matrix with a coherent in-state $|N \ra$ which has $N$  particles on average.

\begin{figure} 
	\centering
	\includegraphics[scale=0.43]{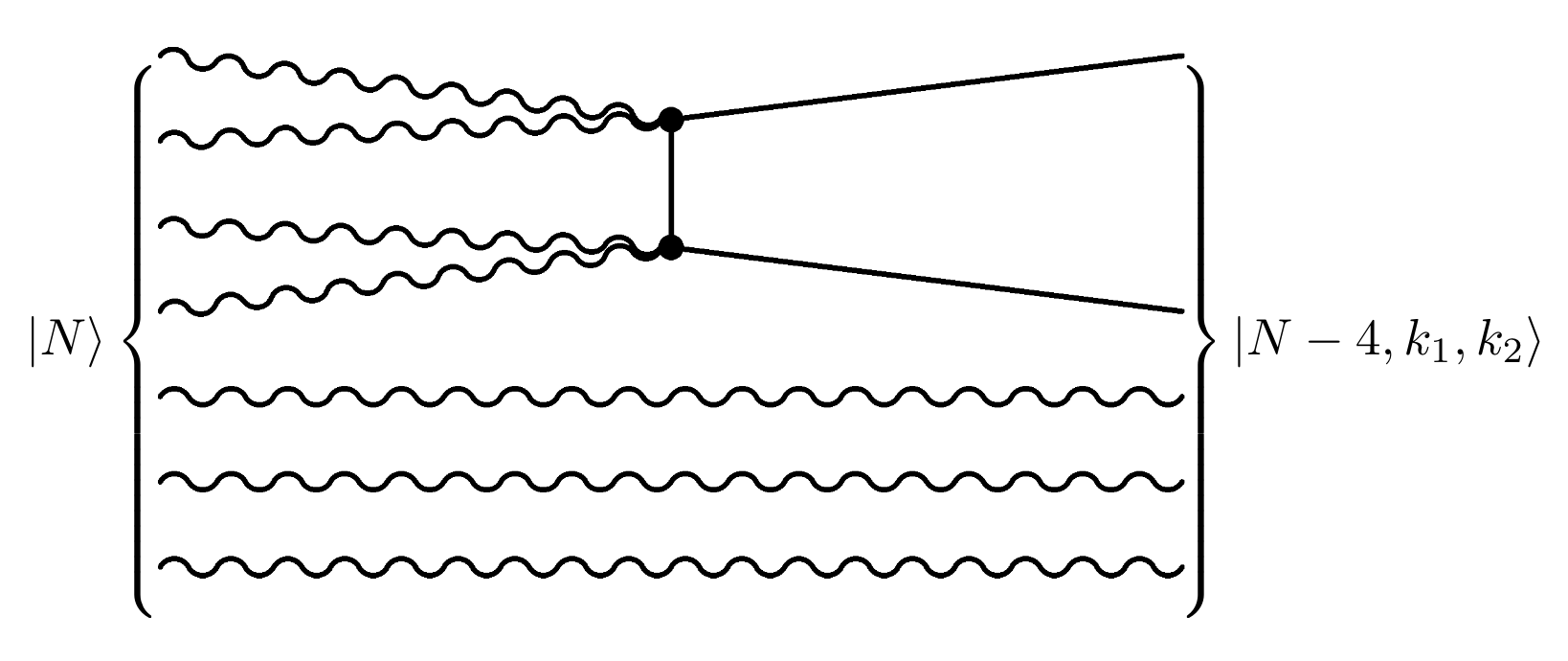} 
	\caption{ Here, the tree-level Feynman diagram associated to the $4\rightarrow 2$ channel is represented. This is the main S-matrix process that leads to the depletion of the condensate. For large $N$, the final state is well approximated by its label (neglecting entanglement).}
		\label{fig:fig2}
\end{figure}

Obviously, there are various quantum channels for the evolution of the system. The most evident ones are the particle number changing processes that lead to the gradual depletion of the condensate, due to the absence of the symmetry protecting against it. Within the theory at hand, the simplest process that leads to the deterioration of the condensate corresponds to  $4\rightarrow 2$ annihilation of the constituents and is depicted in Fig. \ref{fig:fig2}. The rate of depletion through this channel in unit volume is given by \cite{Dvali:2017eba}
\beq
\frac{\Gamma}{V}\sim\left( \lambda \frac{n}{m^3} \right)^4 m^4\,,
\label{ratepv}
\eeq
which is a leading order estimate in the limit of large occupation number.
The quantity appearing in parenthesis is the collective coupling, determined as the multiple of the quantum coupling and the occupation number within the volume of order $m^{-3}$. It is intrinsically a classical quantity that sets the strength of classical nonlinearities. In other words, it would be independent of $\hbar$ if we were to reintroduce it explicitly \cite{Dvali:2017eba}. Using \eqref{ratepv} one can readily obtain the timescale after which an order one fraction of the condensate is expected to be depleted \cite{Dvali:2017eba} (purely through $4\rightarrow 2$)
\beq
t^{-1}_{\rm dep}\sim \lambda \left( \lambda \frac{n}{m^3} \right)^3 m\,.
\eeq
This is precisely the effect captured by the gradual decline in the amplitude of oscillation for the 1-point function, depicted in Fig. \ref{fig:pert}. To convince oneself in this equivalence it suffices to notice that the $\lambda^4$-term of \eqref{eq:1loop}, which is responsible for the decay, contains the required number of $\phi_0$'s for reproducing \eqref{ratepv}.

Let us reiterate that the semi-classical approximation adopted throughout this work thus far is reliable as long as the depletion is negligible, i.e. for $t\ll t_{\rm dep}$. The extrapolation of the dynamics beyond this timescale requires inclusion of higher-order loop effects. As long as we are dealing with leading order quantum corrections to the dynamics, our analysis demonstrates that the elementary S-matrix process responsible for the decline of the 1-point function is the $4\rightarrow 2$ annihilation. As far as the evolution of the state is concerned, on the other hand, one might expect that $2\rightarrow 2$ rescattering of constituents might also lead to the nontrivial dynamics \cite{Dvali:2017eba}. There it was argued that even though such a channel is kinematically prohibited for on-shell particles, it may still proceed due to inevitable off-shell-ness of interacting degrees of freedom. What we have demonstrated in this work is that this effect does not show up in the dynamics of $\la C|\hat{\phi}|C\ra$.

There are, obviously, multi-particle annihilation processes contributing to the depletion as well, e.g. one could have $\delta N> 4$ particles out of total $N$ annihilating into few quanta. It is straightforward to show that the rate of such processes involves higher powers of $[\lambda n/m^3]$ compared to \eqref{ratepv}. Consequently,  even though for small collective couplings the main contributor to the depletion is naturally expected to be the $4\rightarrow 2$ process, for order one collective couplings (and greater) one might naively expect other multi-particle channels to become equally efficient if not dominant. However, as it is demonstrated by our numerical analysis from the previous section, which was carried out for $[\lambda n/m^3]\sim 1$, the full 1-loop dynamics is successfully captured by \eqref{eq:1loop} which terminates at $\lambda^4$-order and thus can only account for $4\rightarrow 2$ channel. It was pointed out to us by Gia Dvali and Lukas Eisemann that there is an S-matrix argument behind this suppression of multi-particle channels \cite{Dvali:2020wqi,Dvali:lukas}.

 \section{Beyond 1-loop}
 
 An important point we would like to discuss further concerns with the choice of the vacuum, around which the coherent state is constructed. The observation was made in \cite{Berezhiani:2020pbv} that, for the purposes of 1-loop computation and to $\lambda^2$-order in semiclassical nonlinearities (i.e. to $\mathcal{O}(\hbar,\lambda^2)$), the vacuum could have been taken to be of the free theory without altering results. In other words, the results were insensitive to the replacement $|\Omega\ra \rightarrow |0\ra$ in \eqref{eq:coherent}. Although it is expected to be legitimate for asymptotic coherent states, the equivalence was counterintuitive for a physical coherent state constructed at a finite time and was believed to be an artefact of the analyzed perturbative order.
 
 In this work, we have extended the analysis to include higher order semiclassical nonlinearities and found $|\Omega\ra \rightarrow |0\ra$ to be inconsequential at 1-loop. We have further compared the two cases including the leading 2-loop corrections, that are $\mathcal{O}(\hbar^2,\lambda^3)$. As a result of the tedious but straightforward computation, we found the results to be identical as long as the construction around the free vacuum is handled in analogy with asymptotic states of the S-matrix formalism. Namely, the 1-point function $\la C_0|\hat{\phi}|C_0\ra (t)$ for
 \beq
 |C_0\ra=e^{-i\int {\rm d}^3 x \left( \phi_{\rm cl}(x)\hat{\pi}(x)-\pi_{\rm cl}(x)\hat{\phi}(x) \right)}|0_{\rm ph}\ra\,
\label{eq:coherent0}
 \eeq
 matches the outcome for \eqref{eq:coherent}; with $|0_{\rm ph}\ra$ being the vacuum of the free Hamiltonian albeit with renormalized mass. The 2-loop mass renormalization that renders the results finite for \eqref{eq:coherent} as well as \eqref{eq:coherent0} is given by
\begin{equation}
m^2_{\text{ph}}=m^2+\left(\frac{\lambda}{2}-\frac{\lambda^2}{2}\int \frac{{\rm d}^3p}{(2\pi)^3(2E_p)^3}\right)\langle\hat{\phi^2}\rangle+\frac{\lambda^2}{3}\int \frac{{\rm d}^3p{\rm d}^3q}{(2\pi)^6(8E_pE_qE_{p+q})}\frac{1}{(E_p+E_q+E_{p+q})}\,.
\label{eq:2loopmass}
\end{equation}
It is straightforward to see that this expression represent a standard subtraction (c.f. Ref.~\cite{ramond1997field}).

Similar to what happens for coupling renormalization at 1-loop, the 2-loop mass renormalization leaves a residual term which, at the level of the equation of motion, is divergent for $t=0$. Although, the solution for the 1-point expectation value remains finite.
  The issues concerning the initial-time singularity are discussed in the following section.

\section{Initial-Time Singularity}
\label{sec:initial}

The appearance of initial-time singularities is usually perceived as a pathology of initial conditions for the fluctuation field. In the literature, these types of singularities have been attributed to the absence of the appropriate counterterms at the initial-time boundary and which can be introduced by the adjustment of the state \cite{Baacke:1999ia,Collins:2005cm}. As long as we do not think of the state of the system in its entirety, this indeed seems to be a perfectly reasonable guiding principle for choosing initial vacua for which no singularities are encountered at any order in perturbation theory. It was demonstrated in \cite{Berezhiani:2020pbv} that this requirement would indicate the inconsistency of unsqueezed coherent states \eqref{eq:coherent} within interacting quantum field theories, unless another resolution of this puzzling behaviour is found. In this work we would like to take a stance aligned with \cite{Berezhiani:2020pbv} and discuss what could possibly provide such an alternative outcome. Namely, considering the elegance of the construction of these states and the success in capturing expected aspects of the dynamics, we would like to entertain the idea that the initial-time singularity is simply an artefact of the perturbative expansion. The obvious motivation for such reasoning is to avoid the premature dismissal of coherent states as members of the physical Hilbert space. This section is devoted to outlining the essential properties of this initial-time singularity.

As shown in \cite{Berezhiani:2020pbv}, the 1-point function in the coherent state has the following undesirable 1-loop behaviour at the initial time
\beq
\lim_{t\rightarrow 0}\Phi(t)\supset \frac{\lambda^2\phi_0^3}{64\pi^2}~ t^2 \cdot{\rm ln}(mt)\,.
\eeq
Although by itself it has a well defined $t\rightarrow 0$ limit, its second time-derivative (i.e. field-acceleration) has a logarithmic divergence. We have extended the 1-loop analysis to higher orders in $\lambda$ and have established that there are no additional 1-loop contributions with similar peculiar initial time behaviour. We have also computed the leading order 2-loop contribution to this singularity. The results can be conveniently summarized in the form of the equation of motion for the 1-point function
\beq
(\partial_t^2+m_{\rm ph}^2)\Phi(t)+\frac{\lambda_{\rm ph}}{3!}\Phi^3(t)+\ldots-\frac{\lambda_{\rm ph}^2}{2}\Phi(t)\phi_0^2\int \frac{{\rm d}^3p}{(2\pi)^3(2E_p)^3}{\rm cos}(2E_pt)\nonumber\\
-\frac{\lambda^2}{3}\phi_0\int \frac{{\rm d}^3p{\rm d}^3q}{(2\pi)^6(8E_pE_qE_{p+q})}\frac{{\rm cos}\left[ (E_p+E_q+E_{p+q})t \right]}{E_p+E_q+E_{p+q}}+\mathcal{O}(\hbar^2,\lambda^3)=0\,,
\label{singeq}
\eeq
where ellipsis stands for manifestly finite terms (including at the initial time) and we have renormalized the parameters where relevant. The last two terms are finite everywhere except at $t=0$, at which moment they diverge; implying the divergence of $\partial_t^2\Phi(t=0)$. An interesting point, however, is that $\Phi(t)$ itself is regular everywhere including $t=0$.

For further clarity let us evaluate \eqref{singeq} at the initial time (since finite quantum corrections depicted by ellipsis vanish)
\beq
&&\partial_t^2\Phi(t=0)+m_{\rm ph}^2\phi_0+\frac{\lambda_{\rm ph}}{3!}\phi_0^3(t)-\frac{\lambda_{\rm ph}^2}{2}\phi_0^3\int \frac{{\rm d}^3p}{(2\pi)^3(2E_p)^3}\nonumber\\
&&~~~~~~~~~~~~~~-\frac{\lambda^2}{3}\phi_0\int \frac{{\rm d}^3p{\rm d}^3q}{(2\pi)^6~8E_pE_qE_{p+q}(E_p+E_q+E_{p+q})}+\mathcal{O}(\hbar^2,\lambda^3)=0\,.
\eeq
It is straightforward to notice that the last two terms are undoing the coupling renormalization and the sunrise contribution to the 2-loop mass renormalization. This implies, for instance, that if the bare coupling is truly infinite then the initial field-acceleration is also infinite. Since we know for a fact that the bare coupling constant is infinite at any given order in perturbation theory, the only possibility for the coherent state in question to avoid $\partial_t^2\Phi(t=0)=\infty$ would be if the bare coupling comes out finite after resumming all infinite contributions. A similar argument should apply to some of the divergent contributions to the mass as well. 

Interestingly, evaluation of the expectation value of Heisenberg's equation, even for a generic coherent state \eqref{eq:coherent}, directly at the initial time yields
\beq
\partial_t^2\Phi(t=0)-\Delta \phi_{\rm cl}+\phi_{\rm cl} \left(m^2+\frac{\lambda Z}{2}\la\Omega|\hat{\phi}^2|\Omega\ra\right)+\frac{\lambda Z}{3!}\phi_{\rm cl}^3=0\,,
\label{acceq}
\eeq
keeping in mind that $\phi_{\rm cl}(x)$ is simply a configuration that determines the coherent state and sets the initial 1-point function $\Phi(t=0)=\phi_{\rm cl}$. Here we have introduced the wavefunction normalization $Z$, as it is nontrivial at 2-loop. It is straightforward to fish out all the counter terms the coherent state is missing at the initial time  from \eqref{acceq}.

Taking into account the comment we have made above, about the initial-time singularities being absent from the $\Phi$ itself while present in its second time-derivative, one could wonder whether it poses a puzzle at all. Even if one were to prefer its avoidance, one could simply start counting time not from the moment when the state is \eqref{eq:coherent} but its evolved version by an infinitesimal time interval $\epsilon$
\beq
|C_{\rm init}\ra=e^{-i H\epsilon}e^{-i\int {\rm d}^3 x \left( \phi_{\rm cl}(x)\hat{\pi}(x)-\pi_{\rm cl}(x)\hat{\phi}(x) \right)}|\Omega\ra\,.
\label{latetimestate}
\eeq

However, the important point we would like to make is that if certain perturbative infinities are not resummed into finite results then the problem with \eqref{eq:coherent} cannot possibly be localized at the initial time. To elucidate the shortcomings, let us go back to the Schr\"odinger picture for a moment. As a result of the time evolution the coherent state \eqref{eq:coherent} begins to loose coherence\footnote{Here we mean coherence in a physical sense, that is connected to the classicality of the state. It, of course, remains to be the coherent state in terms of the general mathematical definition.} gradually. Moreover, the resulting state should have similar (not precisely the same) lack of coherence as the state resulting from evolving \eqref{eq:coherent} backwards in time. The latter, on the other hand, should possess pathologies way before reaching $t=0$, since a well-defined state in renormalizable theory should not be able to evolve into a pathological one. Because of this, one might wonder if there is indeed a pathology that is present in \eqref{eq:coherent} at all times, irrespective of whether it is rewinded forward or backwards, that does not show up in the late time dynamics of correlation functions. Indeed, the expectation value of the Hamiltonian reduces to
\beq
\la C|\hat{H}|C \ra =\int {\rm d}^3x ~Z\left[ \frac{1}{2}\dot{\phi}_{\rm cl}^2+\frac{1}{2}\left(\vec{\nabla}\phi_{\rm cl} \right)^2+\frac{1}{2}\left( m^2+\frac{\lambda Z}{2}\la \Omega|\hat{\phi}^2|\Omega\ra \right)\phi_{\rm cl}^2 +\frac{\lambda Z}{4!}\phi_{\rm cl}^4 \right]\,;
\label{energy}
\eeq
where we have rewritten $\pi_{\rm cl}$ in terms of the initial field-velocity via $\pi_{\rm cl}/Z= \la C| \dot{\hat{\phi}} |C \ra (t=0)\equiv \dot{\phi}_{\rm cl}$; see \cite{Berezhiani:2020pbv} for the derivation. Being a conserved quantity, the pathologies exhibited by \eqref{energy} are possessed by \eqref{latetimestate} as well. Hence, certain perturbative divergencies must resum into finite results\footnote{We would like to thank Otari Sakhelashvili for valuable discussions around this point, through ongoing work on a related subject \cite{Oto}.}, otherwise one would have to accept that the state \eqref{latetimestate} is not a member of the physical Hilbert space, for any $\epsilon$. Examination of \eqref{acceq} and \eqref{energy} lets us conclude that the consistency of the coherent state \eqref{eq:coherent} requires the finiteness of $\lambda$, $Z$ and $\{m^2+\lambda Z/2~\la \Omega|\hat{\phi}^2|\Omega\ra\}$. Notice that the latter contains only bubbly divergencies and is missing some contributions that appear in a standard perturbative mass renormalization; e.g. at 2-loop it is missing the sunrise contribution.

\section{Outlook}
 
 We have analysed certain dynamical aspects of coherent states for massive scalar field with quartic self-interaction. It has been long known that, within the theory at hand, the semi-classical backreaction from quantum fluctuations on the background field itself leads to the gradual deterioration of the amplitude of anharmonic oscillations \cite{baacke1997nonequilibrium}. We have re-examined the process within the coherent state description of the homogeneous condensate and have demonstrated that its depletion is due to annihilation of constituents into relativistic quanta, by linking our findings with the results of the S-matrix analysis of \cite{Dvali:2017eba,Dvali:2017ruz}. In particular, the effect has been established to be dominated by $4\rightarrow 2$ channel at least for moderate collective couplings. The background field method has proven extremely useful in capturing physical effects. It provides a way of analysing perturbative dynamics that avoids spurious secular instabilities that come hand-in-hand with a direct perturbative evaluation of correlation functions \cite{Berezhiani:2020pbv}.

The advantage of knowing a complete form of a quantum state is in the possibility of evaluating any correlation function to any given order in loop expansion. However, certain questions require dynamics of only limited number of correlation functions. The lack of necessity of the entire quantum state is one of the main virtues of this approach, however such a piecewise treatment of the state may obscure certain important aspects. For instance, when one is interested in the leading order quantum corrections to the dynamics of the field configuration it suffices to isolate the dynamics of the 1- and 2-point functions, yielding 1-loop corrected background field and the tree-level dynamics of the 2-point function, which in turn is equivalent to finding the dynamics of the fluctuations around the classical background. It has been long known that certain initial conditions for the semi-classical fluctuations lead to the perturbative back-reaction on the equation of the background field itself that is singular at the initial time. Initial states exhibiting such a behaviour are immediately dismissed within the common lore. Building on the work of \cite{Berezhiani:2020pbv}, we have demonstrated here that the simplest coherent states constructed at a finite instance of time out of well-defined ingredients possess such an unpleasant singularity. Consequently, these are either artefacts of the perturbation theory or the non-squeezed coherent states of the form \eqref{eq:coherent} must be deemed unphysical. For example, we have demonstrated that unless perturbatively divergent bare coupling constant resums to a finite value, the energy of the coherent state corresponding to the localized classical field configuration would be infinite; same applies to the field normalization. Our computation illustrates the advantage of thinking about the state of the system in its entirety as we are able to derive, e.g., the energy of the system without loop expansion.

One may wonder what does the depletion of the background field, defined as the expectation value of the field in the coherent state, imply about the classicality of the system. The significant deviation from the coherence, as increasing number of energetic particles is being produced, definitely indicates that its dynamics is no longer semi-classical. The particle production can be treated as the squeezing of the coherent state to some extent. However, it becomes more nontrivial than that the moment significant energy budget is outsourced from 1-point function to higher-order correlators. One may wonder if the state itself may maintain classicality despite such a loss of coherence. Indeed, if a coherent state were to evolve into, say, a number eigenstate with most of the quanta on a level with macroscopic occupancy then the state itself would be regarded as being close to classical, even though the trajectory that led us there would have been fully quantum. In other words, even if coherence is completely gone, it does not necessarily mean that the state cannot have classical features.

Let us conclude by pointing out that the considered coherent states are parametrized by classical information, i.e. by the initial field configuration $\phi_{\rm cl}(x)$ and $\pi_{\rm cl}(x)$. In reality the state of a system depends on the formation history and may include additional quantum imprints. An example could be the squeezed coherent state, which has the same initial 1-point functions as the simplest states considered here, but has different initial conditions for 2-point functions. Obviously, such modifications would alter the dynamics of not only the 2-point correlation function but the background field as well. In other words, the depletion process we have discussed through the particle production would depend on the origin of the system to some extent. These kind of considerations have been shown to be of utmost importance for black holes \cite{Averin:2016ybl,Averin:2016hhm,Dvali:2018xpy,Dvali:2020wft}. Another important aspect that has been stressed upon in a series of papers \cite{Dvali:2011aa,Dvali:2012wq,Dvali:2012en,Dvali:2013vxa,Dvali:2013eja} concerns with the non-thermal quantum corrections to the black hole spectrum. The point is that the quantum depletion of the background $\Phi$ in \eqref{2pointpsi} affects the spectrum of fluctuations in a nontrivial way, not to mention the contribution from the last two terms which may affect the dynamics at the same order in $\hbar$ as the background depletion. The rigorous application of this point to gravitational systems is technically challenging and will be attempted elsewhere.

\section*{Acknowledgements}

We would like to thank Gia Dvali, Lukas Eisemann and Otari Sakhelashvili for fruitful discussions.


\appendix
\section{1-point function up to $\mathbf{\lambda^4}$}
We begin this appendix by highlighting the steps needed for evaluating the expectation value of a general operator $O[\hat{\phi}]$ over a coherent state $|C\rangle$, using the tools adopted in \cite{Berezhiani:2020pbv}. In particular, the quantity $\langle C|O[\hat{\phi}]|C\rangle$ can be converted into an out-of-time-ordered vacuum expectation value in the interaction picture
\begin{equation}
\langle C|O[\hat{\phi}]|C\rangle=\lim_{T\rightarrow \infty}\frac{\langle 0|U_I(T,0)e^{i\phi_0\int {\rm d}^3x' \hat{\pi}_I(0,x')}U_I(0,t)O[\hat{\phi}_I(t,x)]U_I(t,0)e^{-i\phi_0\int {\rm d}^3x'' \hat{\pi}_I(0,x'')} U_I(0,-T)|0\rangle}{\langle 0|U_I(T,-T)|0\rangle}
\label{eq:evolution2}
\end{equation}
The field operator $\hat{\phi}_I(t,x)$ satisfies the usual ladder expansion
\begin{equation}
\hat{\phi}_I(t,x)=\int \frac{{\rm d}^3p}{(2\pi)^3}\frac{1}{\sqrt{2E_p}}\left(\hat{a}_p e^{ix_\mu p^\mu}+\hat{a}^\dagger_p e^{-ix_\mu p^\mu}\right){\Big\lvert}_{t_0=x-x_0}
\end{equation}
where $t_0$ is a fiducial moment in time where the expansion has been defined with $p_0=\sqrt{m^2+p^2}$. 
Moreover, since all operators on the right-hand side of \eqref{eq:evolution2} are in interaction picture, we introduce the following compact notation 
\begin{equation}
\hat{\phi}_I(t,x)\equiv \phi_{x}\,,\qquad \hat{\phi}_I(t_i,z_i)\equiv\phi_{i}\,.
\end{equation}

It follows from the Baker--Campbell--Hausdorff identity that the operators enclosed by the two exponentials introduced by the presence of a coherent state reduce to 
\begin{equation}
    e^{i\phi_0\int {\rm d}^3x' \hat{\pi}_I(0,x')}O[\phi_x]e^{-i\phi_0\int {\rm d}^3x'' \hat{\pi}_I(0,x'')}=O[\Phi_0(t)+\phi_x]
    \label{eq:shift}
\end{equation}
where $\Phi_0(t)=\phi_0\cos mt$ is the classical solution to the free-equation of motion;  see \cite{Berezhiani:2020pbv}.
Therefore, the inner part of equation \eqref{eq:evolution2} may be calculated dropping the exponents and shifting the fields of the inner operators by the solution of the free equation of motion.

As an explicit example, we apply $\eqref{eq:evolution2}$ to calculate the 1-point function up to $\lambda^3$ corrections. The generalization to other correlators is straightforward.
Concerning the theory at hand, operators $U_I(T,0)$ and $U_I(0,-T)$ that connect the non-interacting vacuum to the true vacuum may be dropped, since they will not contribute to the 1-point function at 1-loop. This simplifies \eqref{eq:evolution} to:
\begin{equation}
\langle C|\hat{\phi}|C\rangle={\langle 0|e^{i\phi_0\int {\rm d}^3x' \hat{\pi}_I(0,x')}U_I(0,t)\phi_x U_I(t,0)e^{-i\phi_0\int {\rm d}^3x' \hat{\pi}_I(0,x')} |0\rangle}+\mathcal{O}(\hbar^2)
\label{eq:evolution}
\end{equation}
In other words, at 1-loop, the interacting vacuum of the theory may be identified with its asymptotic vacuum. 
Now, we expand the evolution operators in $\left\{U_I(0,t)\phi_x U_I(t,0)\right\}$ up to the third order in $\lambda$ and apply the identity \eqref{eq:shift}, that removes the exponentials and shifts all $\hat{\phi}$ fields by $\Phi_0(t)$. Once the contractions are performed and the propagators are evaluated in momentum space, the final expression for the 1-loop contributions to $\langle C|\hat{\phi}|C\rangle$ up to $\lambda^3$ is found.
We redirect the reader to \cite{Berezhiani:2020pbv} for a detailed evaluation of those steps we just described, up to $\lambda^2$. The generalization to $\lambda^3$ corrections is straightforward (even though more tedious).

The $\lambda$ and  $\lambda^2$ contributions are
\begin{flalign}
\langle C|\hat{\phi}|C\rangle_{\lambda+\lambda^2}=\Phi^\text{cl}_{\lambda^2}(t)+\frac{\lambda^2}{2}\int \frac{{\rm d}^3p}{(2\pi)^ 3(2E_p)^2}\Phi_0(t_1)\Phi^2_0(t_2)\sin 2E_p(t_1-t_2)\frac{\sin(m(t-t_1))}{m}\,,
\label{eq:quadratic}
\end{flalign}
where $\Phi^\text{cl}_{\lambda^2}(t)=\Phi_0(t)+\Phi_1(t)+\Phi_2(t)$ is the classical solution to the equation of motion up to $\lambda^2$, with $\Phi_n(t)$ denoting $\mathcal{O}(\lambda^n)$ classical contributions, while the integral represents the first non-trivial quantum correction to the one-point function. Even though it is manifestly divergent, that contribution can be made finite imposing coupling renormalization.

Then we have $\lambda^3$ terms, with the classical part 
\begin{flalign}
   \langle C&|\phi|C\rangle_{\lambda^3}^{\text{cl}}=\frac{\lambda ^3 \phi_0 ^7}{4!^3 2^9m^6} \biggr(\left(1080 m^2 t^2-547\right) \cos (m t)+\left(594-648 m^2 t^2\right) \cos (3 m t)\nonumber\\&+24 m t \sin (m t) \left(12 m^2 t^2+88 \cos (2 m t)-5 \cos (4 m t)-25\right)-48 \cos (5 m t)+\cos (7 m t)\biggr)\,,
   \label{eq:lambda3cl}
\end{flalign}
which, matches the $\lambda^3$ correction to the classical solution of the equation of motion for a scalar oscillator endowed by a quartic potential, as expected.
Finally, order $\hbar$ quantum contributions at $\lambda^3$ order are
{\small
\begin{align}
-\frac{\lambda^3}{2}\int \frac{{\rm d}^3p}{\left(2\pi\right)^3\left(2E_p\right)^3} \int_0^t{\rm d}t_1\int_{0}^{t_1} {\rm d}t_2\int_{0}^{t_2} {\rm d}t_3 \Phi_0(t_1)\Phi_0^2(t_2)\Phi_0^2(t_3)\frac{\sin m(t-t_1)}{m}\times\nonumber\\\times\left(\cos(2E_p(t_2-t_3))-\cos(2E_p(t_1-t_3))\right)\nonumber\\
+\frac{\lambda^2}{2}\int \frac{{\rm d}^3p}{(2E_p)^2(2\pi)^3}\int_0^t {\rm d}t_1\int_0^{t_1}{\rm d}t_2\biggr[\left(\Phi_1(t_1)\Phi^2_0(t_2)+2\Phi_0(t_1)\Phi_0(t_2)\Phi_1(t_2)\right)\sin(2E_p(t_1-t_2)\nonumber\\+\frac{\lambda}{2} \int_0^{t_2}{\rm d}t_3\Phi_0^2(t_1)\Phi_0(t_2)\Phi_0^2(t_3)\sin(2E_p(t_2-t_3))\frac{\sin{m(t_1-t_2)}}{m}\biggr]\frac{\sin m(t-t_1)}{m}\,.
\label{eq:cube}
\end{align}}
It is interesting to see that the third line may be combined with the $\lambda^2$ one-loop term of \eqref{eq:quadratic} resulting in
{\small
\begin{flalign}
\frac{\lambda^2}{2}\int \frac{{\rm d}^3p}{(2E_p)^2(2\pi)^3}\int_0^t {\rm d}t_1\int_0^{t_1}{\rm d}t_2&\Phi_\lambda(t_1)\Phi_\lambda^2(t_2)\sin(2E_p(t_1-t_2))\frac{\sin m(t-t_1)}{m}\,,
\label{eq:nonlin}
\end{flalign}}
while the fourth line combines with all the classical terms up to $\lambda^2$ into:
\begin{flalign}
\frac{\lambda}{3!}\int_0^t {\rm d}t_1 \Phi_{\lambda^2}^3(t_1)\frac{\sin(m(t-t_1))}{m}\,.
\label{eq:cubic}
\end{flalign}
Here, $\Phi_{\lambda}$ and $\Phi_{\lambda^2}$ are the 1-point expectation value truncated at, respectively, the $\lambda$ and $\lambda^2$ order.

Therefore,  Eq.~\eqref{eq:nonlin} could have been obtained from the $\lambda^2$ contribution replacing $\Phi_0(t)$ with the full classical solution to the equation of motion. This behaviour is a consequence of the fact that two classes of one-loop quantum corrections arise from the perturbative expansion \eqref{eq:evolution}. The first class, the \textit{irreducible} terms, are contributions that cannot be resummed with previous order terms. For example, the irreducible $\lambda^2$ term is given by the integral in \eqref{eq:quadratic} while for $\lambda^3$ we have the first two lines of \eqref{eq:cube}. The second class of terms is given by \textit{reducible} contributions, which are terms which may be resummed with the irreducible contributions appearing at the previous orders, as shown for \eqref{eq:nonlin} and \eqref{eq:cubic}. It is straightforward to see that the solution of \eqref{eq:1loopeom} up to $\lambda^3$ contains both the reducible and irreducible $\lambda^3$ contributions we have found in this section. In particular, \eqref{eq:cubic} is the part of the solution of  \eqref{eq:1loopeom} sourced by the cubic term of the e.o.m, while \eqref{eq:nonlin} is sourced by the classical non-linearities of the background insertions in the $\lambda^2$ and 1-loop term.

Finally, let us comment on $\lambda^4$ terms. According to the statement given above, we may obtain the reducible $\lambda^4$ contributions by correcting the condensate insertions in \eqref{eq:cubic} from $\Phi_{\lambda^2}(t_1)$ to $\Phi_{\lambda^3}(t_1)$ (and in a similar way in \eqref{eq:nonlin}) and then expanding up to $\lambda^4$ and $\hbar$.  What we are missing are the $\lambda^4$ irreducible terms, which are responsible for the depletion of the one-point function. Diagrammatically those contributions look like an on-shell propagator entering a loop with four vertices. Therefore, we may find them expanding Eq. \eqref{eq:evolution} up to $\lambda^4$ and fishing out contractions that look like
\begin{equation}
\frac{\lambda^4}{2^7}\Phi_0(t_1)\Phi_0^2(t_2)\Phi_0^2(t_3)\Phi_0^2(t_4)\langle \phi_{z_1}^2\phi_{z_2}^2\phi_{z_3}^2\phi_{z_4}^2\rangle+...
\end{equation}
If one collects those terms and express propagators in momentum space, the $\lambda^4$ irreducible contribution will reduce to
{\small
\begin{flalign}
\frac{\lambda^4}{8}\int&\frac{{\rm d}^3p}{(2\pi)^3 (2E_p)^4}\int_0^t {\rm d}t_1\int_0^{t_1}{\rm d}t_2\int_0^{t_2}{\rm d}t_3\int_0^{t_3}{\rm d}t_4\Phi_0(t_1)\Phi_0^2(t_2)\Phi_0^2(t_3)\Phi_0^2(t_4)\biggr(2\sin 2E_p(t_1-t_4)\nonumber\\&-2\sin 2E_p(t_2-t_4)+\sin(2E_p(t_1-t_2+t_3-t_4)+\sin(2E_p(t_1-t_2-t_3+t_4))\biggr)\frac{\sin(m(t-t_1))}{m}\,,
\end{flalign}}
which matches the $\lambda^4$ term obtained from solving the equation of motion \eqref{eq:1loop}.

\section{Two-loop mass renormalization}
In Ref.~\cite{Berezhiani:2020pbv}, it was verified how the 1-loop mass and coupling prescriptions commonly adopted in the S-matrix formalism are the same ones regularizing the coherent state time-evolution. In this appendix, we check if this statement holds at the 2-loop order and, in particular, how the initial-time singularity arises at higher order in $\hbar$. In order to verify these two points, let us study again the equation of motion for the 1-point function. However, because we are not limiting our analysis only to 1-loop contributions, the full e.o.m. must be considered, namely
\begin{equation}
\left( -\Box+m^2 \right)\Phi(x,t)+\frac{\lambda}{3!}\langle C|\hat{\phi}^3(x,t)|C\rangle=0\, .
\label{eq:2loop}
\end{equation}
After a straightforward calculation of $\langle C|\hat{\phi}^3(x,t)|C\rangle$ up to $\mathcal{O}(\lambda)$ and 2-loop terms, Eq.~\eqref{eq:2loop} reduces to
\begin{flalign}
\nonumber\left(-\square+m^2\right)\Phi(t)&+\frac{\lambda}{3!}\Phi^3(t)+\frac{\lambda}{2}\langle\phi^2\rangle\Phi(t)-\frac{\lambda^2}{2}\langle\phi^2\rangle\Phi(t)\int \frac{{\rm d}^3p}{(2\pi)^3(2E_p)^3}+\\&\nonumber-\frac{\lambda\phi_0}{8m^2}\langle\phi^2\rangle^2 mt \sin m -\frac{\lambda^2}{2}\Phi(t)\int \frac{{\rm d}^3p}{(2\pi)^3(2E_p)^2}\int_0^t{\rm d}t_1\Phi^2(t_1)\sin2E_p(t_1-t)+\nonumber\\&+\frac{\lambda^2}{3}\int \frac{{\rm d}^3p{\rm d}^3q}{(2\pi)^6(8E_pE_qE_{p+q})}\int_0^t\Phi(t_1)\sin[(E_p+E_q+E_{p+q})(t-t_1)]=0
\label{eq:2loopexplicit}
\end{flalign}
Using the result of the previous appendix, we replaced the free solution $\Phi_0(t)$ by the full solution $\Phi(t)$. 

Again, solving the 2-loop equation of motion gives the same result obtained perturbatively in \cite{Berezhiani:2020pbv}.
The divergent parts of the sixth term could be reabsorbed imposing the 1-loop coupling renormalization, since the first 2-loop contribution to the coupling enters at order $\lambda^3$. 

Concerning the mass, the condition 
\begin{equation}
m^2_{\text{ph}}=m^2+\left(\frac{\lambda}{2}-\frac{\lambda^2}{2}\int \frac{{\rm d}^3p}{(2\pi)^3(2E_p)^3}\right)\langle\hat{\phi}^2\rangle+\frac{\lambda^2}{3}\int \frac{{\rm d}^3p{\rm d}^3q}{(2\pi)^6(8E_pE_qE_{p+q})}\frac{1}{(E_p+E_q+E_{p+q})}\,,
\label{eq:2loopmassAppendix}
\end{equation}
absorbs all the divergences. Inverted this relation, the expression for the bare mass in terms of physical quantities readily follows
\begin{equation}
 m^2=m^2_{\text{ph}}-\left(\frac{\lambda_\text{ph}}{2}+\frac{3\lambda^2_\text{ph}}{2}\int \frac{{\rm d}^3p}{(2\pi)^3(2E_p)^3}\right)\langle\hat{\phi}^2\rangle_{ph}-\frac{\lambda^2}{3}\int \frac{{\rm d}^3p{\rm d}^3q}{(2\pi)^6(8E_pE_qE_{p+q})}\frac{1}{(E_p+E_q+E_{p+q})}\,,
\end{equation}
where all the $E_p$ factors have to be intended in terms of physical masses. It is straightforward to verify that these expressions contain standard divergencies encountered in S-matrix computations (c.f. Ref.~\cite{ramond1997field}).


\printbibliography

\end{document}